\documentclass[%
 notitlepage,
 twocolumn,
nofootinbib,
 amsmath,amssymb,
 aps,
 pra,
]{revtex4-1}

\usepackage{graphicx}
\usepackage{physics}
\usepackage{dcolumn}
\usepackage{bm}
\usepackage{bbold}
\usepackage{soul}

\usepackage{color}
\begin{document}


\title{Producing and storing spin-squeezed states and Greenberger-Horne-Zeilinger states in a one-dimensional optical lattice}

\author{Marcin P\l{}odzie\'n}
\affiliation{International Research Centre MagTop, Institute of
  Physics, Polish Academy of Sciences, Aleja Lotnik\'ow 32/46, PL-02668 Warsaw, Poland}

\author{Maciej Ko\'scielski, Emilia Witkowska}
\affiliation{Institute of Physics, Polish Academy of Sciences, Aleja Lotnik\'ow 32/46, PL-02668 Warsaw, Poland}

\author{Alice Sinatra}
\affiliation{Laboratoire Kastler Brossel, ENS-Universit\'e PSL, CNRS, Sorbonne Universit\'e, Coll\`ege de France, 24 rue Lhomond, 75005 Paris, France}

\date{\today}

\begin{abstract}

We study the dynamical generation and storage of spin squeezed states, as well as more entangled states up to macroscopic superpositions, in a system composed by a few ultra-cold atoms trapped in a one-dimensional optical lattice. The system, initially in the superfluid phase with each atom in a superposition of two internal states, is first dynamically entangled by atom-atom interactions then adiabatically brought to the Mott-insulator phase with one atom per site where the quantum correlations are stored. Exact numerical diagonalization allows us to explore the structure of the stored states by looking at various correlation functions, on site and between different sites, both at zero temperature and at finite temperature, as it could be done in an experiment with a quantum-gas microscope. 
\end{abstract}

\maketitle

\section{Introduction}

Ultra-cold atoms are widely recognized as a platform for fundamental tests of quantum mechanics, quantum information and quantum networks~\cite{doi:10.1080/00018730701223200,RevModPhys.80.885,RevModPhys.86.153,Bloch2012,Gross995}, as well as sensors in well established applications like atomic clocks~\cite{RevModPhys.87.637, Campbell90, PhysRevLett.123.123401}. In some implementations, non-trivial correlations among the atoms in the form of squeezing and entanglement are directly profitable. In addition, recent developments in experimental techniques allow to produce and control ultra-cold systems in the few-body regime with a good precision on the atom number \cite{Serwane336,Wenz457,PhysRevLett.114.080402,PhysRevLett.108.075303}. These experimental developments in ultra-cold few body systems boost theoretical studies \cite{Sowi_ski_2019} indicating new possibilities for thermometry \cite{PhysRevA.97.063619} and quantum simulations  \cite{PhysRevB.100.041116,Chatterjee_2019}.

\begin{figure}[tb!]
	\centering
	\includegraphics[width=0.9\linewidth]{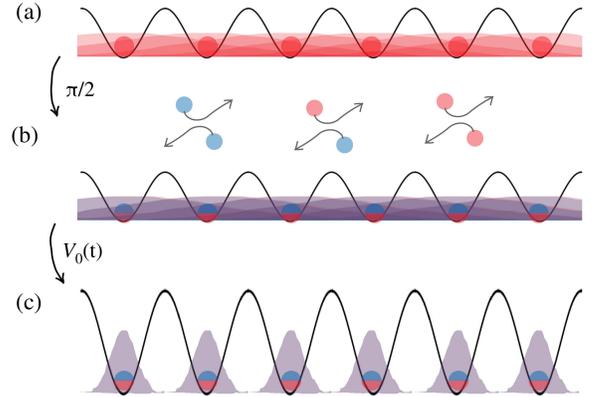}
	\caption{(a) Initially, ultra-cold atoms in an internal state $a$ are prepared in the superfluid phase in a shallow optical lattice with unit filling. (b) At time $t=0$, a $\pi/2$-pulse puts each atom in a coherent superposition of two internal states $a$ and $b$. 
	Immediately after the pulse, while the system in the superfluid phase, binary atomic interactions between atoms start the generation of quantum correlations in the system. (c) The lattice height is gradually increased, in such a way that the system undergoes the Mott-insulator transition with one atom per site at the ``best time", to store in the lattice either the best squeezing or a GHZ state.}
	\label{fig:fig0}
\end{figure}

In a previous work~\cite{Kajtoch_2018} we proposed a method for the dynamical generation of 
spin-squeezing and its storage in an optical lattice with unit filling, see Fig.\ref{fig:fig0}. The scheme makes use of interactions
in a Bose-Einstein condensate with two internal states. The atoms are initially in the superfluid phase delocalized over the entire lattice and in a single internal state. As soon as the atoms are put in a superposition of two internal states by an electromagnetic pulse, atomic interactions dynamically generate non-trivial correlations and spin-squeezing~\cite{nature08988,nature08919}.
The lattice depth is then adiabatically increased.  As the system approaches the Mott-insulator phase, the spin dynamics slows down to stop completely at the transition, while the spin-squeezing at its best and the quantum correlations survive and are stored deeply in the Mott insulating phase. 
The work~\cite{Kajtoch_2018}, that was for a large number of atoms $N \sim 10^5$ and a three-dimensional lattice, was relying on two main approximations: (i) the hypothesis of adiabaticity while ramping the optical lattice in the two-component interacting condensate, and (ii) the Gutzwiller approximation to calculate the ground state energy of the system for different lattice heights.

In the present paper, we deal with a few-bosons in a one-dimensional optical lattice. Within a full representation of the system Hamiltonian~\cite{Zhang_2010}, we study its exact state and dynamics. This allows us to confirm the validity of the scheme beyond the approximations we used in ~\cite{Kajtoch_2018}.
The advantage of the few-body system over the macroscopic number of atoms is furthermore the possibility to access and fully characterize the quantum state stored in Mott-insulator phase,
including the excitation generated by the system manipulations (the initial pulse and the dynamic ramping of the lattice) and by thermal fluctuations. 
From the experimental point of view, the on-site and inter-site correlation functions that we calculate would be accessible in an experiment with a quantum gas microscope~\cite{Bakr547, Weitenberg2011}.

After introducing in Section \ref{sec1} the system Hamiltonian and parameters,
the following Section~\ref{sec:2MM} is devoted to an analytical one-axis twisting model, issued from the zero-momentum part of the two-component (two internal states) Bose-Hubbard Hamiltonian, that catches the main features of the dynamics. By testing the model against exact numerical simulations, we prove its validity for a large enough condensed fraction. One useful result, predicted by the simple model and confirmed by the simulations, is that the squeezing dynamics is faster in a shallow lattice with respect to the homogeneous system. This acceleration is directly related the confinement of the atoms causing an increase of the interaction energy and a consequent reduction of the squeezing time scales. 

In Section  \ref{sec:structure} we study the structure of the states stored in the Mott phase that turns out to be non trivial. 
With respect to what we shall call an ``ideal Mott-insulator phase-state", that has exactly 
one atom per site, each atom being in the same coherent superposition of two internal states, the stored entangled state shows additional phase factors that play a crucial role in determining its properties. We discuss the possibility of storing either a gaussian spin-squeezed state or a maximally entangled Greenberger-Horne-Zeilinger state~\cite{GHZFirst, doi:10.1119/1.16243}, superposition of two ideal Mott-insulator phase-states with different phases, provided that the ramp time is properly chosen. The challenge of generating experimentally a Greenberger-Horne-Zeilinger state has been already taken-up~in several platforms, see e.g.~\cite{Sackett2000, PhysRevLett.92.087902, PhysRevLett.91.180401,PhysRevLett.106.130506, PhysRevLett.82.1345, PhysRevLett.117.210502, PhysRevLett.119.180511}. 
The strength of the method we propose here relies in its relative simplicity, and in the possibility to scale it up to a larger system size, by using a larger lattice and by going from one to three spatial dimensions.

Finally, in Section~\ref{sec:T}, we discuss the role of a finite temperature resulting in particle-hole excitations in the Mott phase that can lead to double occupation of certain lattice sites. We show that it is nevertheless possible to maintain squeezing in the Mott phase at low temperature, when the probability for particle-hole excitation is weak. Concluding remarks and a summary are given in Section \ref{sec:conclusion}.

\section{Considered system and model Hamiltonian}\label{sec1}

In our study, we use the Bose-Hubbard model to describe a system composed of a few ultra-cold bosonic atoms in an optical lattice potential. In the following we discuss the model, parameters that we used and the protocol for dynamical generation of correlations among the atoms.

\subsection{Bose-Hubbard Hamiltonian}
We consider an ultra-cold gas composed of a few bosonic atoms with repulsive interactions in the two internal states 
$a$ and $b$, loaded in a three-dimensional optical lattice or optical tweezers potential. We assume that tunneling is possible only along the $x$ direction while transversely the atoms are in a localized wave-function that is approximated by a Gaussian with a characteristic length $L_\perp$. The optical lattice potential in the $x$-direction is $V(x)=V_{0} \sin^{2}(k x)$, where $V_{0}$ is the lattice height, $k = 2\pi/\lambda$ is the wave number, and $d=\lambda/2$ is the spatial period of the lattice. In the absence of the lattice, the system is homogeneous, the atoms being confined in a flat-bottom potential \cite{PhysRevLett.110.200406}. After reduction of the perpendicular directions, the effective one-dimensional system Hamiltonian reads
\begin{eqnarray}
\hat{H} &=& \sum_{\sigma=a,b} \int dx \left(
\hat{\Psi}_\sigma^\dagger(x) \hat{h} \hat{\Psi}_\sigma(x) + 
\frac{g_{\sigma \sigma}}{2} 
\hat{\Psi}_\sigma^{2\dagger}(x) \hat{\Psi}_\sigma^{2}(x) \right) \nonumber \\
&+& g_{ab} \int dx \: \hat{\Psi}_a^\dagger(x)\hat{\Psi}_b^\dagger(x)\hat{\Psi}_b(x)\hat{\Psi}_a(x)
\label{eq:Hamiltonian}
\end{eqnarray}
with the single-particle Hamiltonian
\begin{equation}\label{eq:single_particle_hamiltonian}
\hat{h} = -\frac{\hbar^2}{2 m} \frac{d^2}{dx^2} + V(x).
\end{equation}
The interaction coefficients $g_{\sigma \sigma'}=g^{3D}_{\sigma \sigma'}/(2 \pi L_\perp^2)$ are determined by the transverse confinement length $L_\perp$ and by the coupling constants in three dimensions $g^{3D}_{\sigma \sigma'}=4\pi \hbar^2 a_{\sigma \sigma'}/m$, where $a_{\sigma \sigma'}$ is the $s$-wave scattering length for one atom in the state $\sigma$ and one in the state $\sigma'$, $m$ is the atomic mass and $\hbar$ is the Planck constant. We assume repulsive interactions between atoms in the two states, with $a \leftrightarrow b$ symmetry leading to $g_{aa} = g_{bb}$, and an adjustable inter-species coupling $g_{ab}$, restricting to the phase-mixing regime~: $g_{ab}<g_{aa}$ \cite{PhysRevLett.81.5718}. 

The system (\ref{eq:Hamiltonian}) in the lowest energy band is conveniently considered in the basis of Wannier functions $w(x-x_i,t)$ localized around the lattice sites, where $x_i$ denotes position of the $i$-th site~\cite{RevModPhys.80.885}. 
In the tight-binding limit when the lattice height is larger than the recoil energy $E_R=\hbar^2k^2/(2m)$ and the Wannier functions are well localized around each lattice site, the tunneling and interactions terms fall off rapidly with the distance $|x_i - x_j|$, leading to the Bose-Hubbard model
\begin{align}\label{eq:BHM}
\hat{\mathcal{H}}_{\rm BH} &= - J \sum\limits_{i, j=i\pm 1} \left(\hat{a}_{i}^{\dagger}\hat{a}_{j} + \hat{b}_{i}^{\dagger}\hat{b}_{j}\right) + \frac{U_{aa}}{2}\sum\limits_{i} \hat{n}^a_i (\hat{n}^a_i -1) \nonumber\\
& + \frac{U_{bb}}{2}\sum\limits_{i} \hat{n}^b_i (\hat{n}^b_i -1) 
+ U_{ab} \sum \limits_{i} \hat{n}^a_i\hat{n}^b_i ,
\end{align}
where $J$ and $U_{\sigma \sigma'}$ are the tunneling and interaction parameters.
$\hat{a}_i$ ($\hat{b}_i$) is the annihilation operator of an atom in internal state $a$ ($b$) in the $i$-th site of the lattice, and $\hat{n}^a_i=\hat{a}_{i}^{\dagger}\hat{a}_{i}$, $\hat{n}^b_i=\hat{b}_{i}^{\dagger}\hat{b}_{i}$ are the corresponding number operators. 
The Bose-Hubbard Hamiltonian commutes with the total number of particles in each component
$\comm{\hat{\mathcal{H}}_{\rm BH}}{\hat{N}_\sigma}=0$, where $\hat{N}_\sigma=\sum_i \hat{n}_{\sigma, i}$ with $\sigma = a,b$, but it does not commute with the occupation numbers $\hat{n}_{a, i},\, \hat{n}_{b, i}$ of the $i$-th site, due to the presence of the hoping terms. We address here the case in which the total filling is commensurate with the lattice.
In the repulsive interaction regime we consider, a transition from a superfluid to a Mott-insulator phase is expected in the two-component system, as in the usual case of a single species~\cite{Bakr547}; the magnetic order associated with the pseudospin degrees of freedom (boson components) being a further feature in the ground state~\cite{Altman_2003}.
We briefly present the construction of the Bose-Hubbard Hamiltonian (\ref{eq:BHM}) from the more general one 
(\ref{eq:Hamiltonian}) in Appendix~\ref{app:BHmodel}. The basic properties of the tunneling and interaction parameters are 
summarized in Appendix~\ref{app:JandU}, where we also calculate for different lattice heights the couplings appearing in (\ref{eq:Hamiltonian}) that are neglected in (\ref{eq:BHM}).

\subsection{Procedure and parameters used in the simulations}
Throughout the paper, we will consider two situations: (i) the case of a static lattice with a time independent height $V_0$, and (ii) the dynamical case where the lattice is raised with a linear ramp in a time $\tau$
\begin{equation}
V_{0}(t) = V_{\rm ini} + (V_{\rm fin} - V_{\rm ini})\frac{t}{\tau},
\label{eq:ramp1}
\end{equation}
from an initial value $V_{\rm ini}$ where the system is in the superfluid regime to a final value $V_{\rm fin}$ in the Mott-insulator regime. The ramp is adjusted in such a way that a particular state (a maximally spin-squeezed state or a more entangled state produced by dynamics) is stored in the Mott phase as the system crosses the Mott transition. Note that in this case $J(t)$ and $U(t)$ are functions of time. 
\begin{figure}[tb!]
\centering
\includegraphics[width=.8\linewidth]{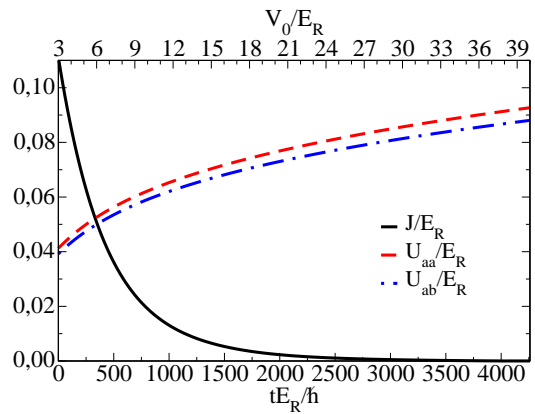}
\caption{Values of the tunneling $J$ (\ref{eq:tunneling}) and interaction $U_{aa}$, and $U_{ab}$, (\ref{eq:u_definition}) parameters across the
lattice ramp (\ref{eq:ramp1}) used in simulations for $N=6$ atoms. Here, $V_{\rm ini}/E_R=3$, $V_{\rm fin}/E_R=40$ and $\tau E_R/\hbar=4255$.}
\label{fig:fig30}
\end{figure}
Using the Bose-Hubbard Hamiltonian~(\ref{eq:BHM}) and periodic boundary conditions, 
we perform exact numerical calculations for a few atoms, concentrating on the case $N=M=6$, where $N$ is the number of atoms and $M$ the number of lattice sites. For the simulations we use the Fock basis $|n\rangle=|\{n_i^a \} , \{ n_i^b\} \rangle$, where $\{n_i^{a(b)} \}$ denotes a configuration of occupation numbers of the $M$ lattice sites for atoms in the internal state $a~(b)$. According to the commutation relations $\comm{\hat{\mathcal{H}}_{\rm BH}}{\hat{N}_\sigma}=0$, 
the Hilbert space can be written as a direct sum of subspaces with a fixed number of atoms in each internal state
$\kappa=\kappa_{N,0} \oplus \kappa_{N-1,1} \oplus \kappa_{N-2,2} \cdots \kappa_{0,N}$,
where $\kappa_{N_a,N_b}$ denotes the subspace with $N_a$ atoms in the state $a$ and $N_b$ in the state $b$. The dimension $d_{N_a, N_b}$ of each subspace depends on $N_{a}$, $N_{b}=N-N_a$ and on the number of lattice sites $M$. One has $\kappa_{N_a, N_b}=\binom{N_a+M-1}{N_a}  \binom{N_b+M-1}{N_b}$.
A numerical method for the Fock state basis generation is described in \cite{Zhang_2010}. Operators $\hat{\mathcal{O}}$ are represented by matrices $\langle n'| \hat{\mathcal{O}}|n\rangle$, and pure states by vectors whose components are the coefficients of the decomposition in the basis.

We consider a spin coherent state as the initial state for the squeezing evolution.
In order to prepare it numerically, we assume that initially all the atoms are in the internal state $a$ and, in the case of zero temperature, in the ground state of $H_{\rm BH}$ for the initial values of~$J$~and~$U$:
\begin{equation}
\ket{\varPsi_0(0^-)} =
\sum_{\{n^{a}_i\} \atop \sum_i n_i^a=N} \alpha_{\{n^{a}_i\}}\ket{\{n^{a}_i\}, \{0\}} ,
\end{equation}
where the sum runs over all the possible configurations, and the coefficients $\alpha_{\{n^{a}_i\}}$ are calculated numerically. A $\pi/2$ pulse, equivalent to a rotation of the state around the $y$-axis through angle $\pi/2$, is applied to create the spin coherent state~\cite{PhysRevA.6.2211}
\begin{equation}\label{eq:initial}
    |\varPsi(0^+)\rangle = e^{-i\hat{S}_y\pi/2}|\varPsi_0(0^-)\rangle.
\end{equation}
When the temperature is nonzero, the initial state after rotation is given by the density matrix:
\begin{equation}\label{eq:rhoT}
    \hat{\rho}_{T}(0^+)= e^{-i \hat{S}_y \pi/2}\hat{\rho}_T(0^-) e^{i \hat{S}_y \pi/2} ,
\end{equation}
with
\begin{equation}
    \hat{\rho}_T(0^-)=\frac{1}{Z}\sum_n e^{-E_n/k_B T} |\Psi_n(0^-)\rangle \langle\Psi_n(0^-)|
\end{equation}
where $E_n$ are eigenenergies of the initial Hamiltonian with $\sum_i n_i^a=N$ and $|\Psi_n (0^-)\rangle$  are the corresponding eigenvectors, $k_B$ is the Boltzman constant, $T$ the temperature and $Z$ the normalization. 
The initial state is then evolved according to the Schr\"odinger equation 
$i\hbar \partial_t|\varPsi(t)\rangle = {\cal \hat{ H}}_{\rm BH}|\varPsi(t)\rangle$ 
for the zero temperature case, or the von Neumann equation $i \hbar \partial_t{\hat{\rho}}_{T}(t) = [ \hat{\mathcal{H}}_{\rm BH}, \hat{\rho}_{T}(t)]$ for the nonzero temperature case. Both equations were solved numerically using the Runge-Kutta method.

In Fig.~\ref{fig:fig30} we show the values of the tunneling parameter $J(t)$ and the interaction parameter $U_{aa}(t)$ through the considered ramp (\ref{eq:ramp1}) optimized to produce and store a spin-squeezed state with $N=6$ atoms.  $J(t)$ and $U_{aa}(t)$ were calculated numerically according to the definitions (\ref{eq:tunneling}-\ref{eq:u_definition}) using the Wannier functions, as detailed in Appendix~\ref{app:BHmodel}.
The numerical values of parameters in all the simulations, are typical for hyperfine states in alkali atoms
\footnote{In an experiment, one could use for example ${}^{87}$Rb atoms in states $|F=1,m_F=1\rangle$ and $|F=2,m_F=-1\rangle$ where the interspecies scattering length $a_{ab}$ is tuned by means of a Feshbach resonance \cite{nature08919},  or the states $|F=1,m_F=-1\rangle$ and $|F=2,m_F=-2\rangle$  where the interspecies interaction can be tuned by slightly shifting the optical lattices for the two components \cite{PhysRevLett.82.1975}.}.
We took $a_a=100.4 a_B$, $a_b = a_a$, and $a_{ab} = 95 a_a$, where $a_B = 5.29\times 10^{-11}$m is Bohr radius, the lattice constant $d = \lambda/2=431$nm, the ratio $(a_a+a_b-2a_{ab})/d=1.25\times 10^{-3}$, $a_a/d=0.012$. 
The recoil energy value is $E_R = 2.04\times10^{-30}$~J and the characteristic time scale $t_0 =\hbar/E_R= 51.5\mu$s. We also assume that $L_\perp=d/\sqrt{2\pi}$.

\subsection{Spin squeezing parameter}\label{sec2}

We quantify the level of spin-squeezing by the parameter~\cite{PhysRevA.46.R6797, PhysRevA.50.67}
\begin{equation}\label{eq:spin_squeezing}
\xi^2 = \frac{N  \Delta^2 \hat{S}_{\perp\, \rm min}}{\langle S\rangle^2},
\end{equation}
where $\langle S\rangle$ is the length of the mean collective spin and $\Delta^2 \hat{S}_{\perp\, \rm min}$ is the minimal variance of the spin orthogonally to the mean spin direction. In the multimode case of atoms in an optical lattice, collective spin operators are defined as a sum of local spin operators
\begin{eqnarray}\label{eq:spinoperators}
\hat{S}_x&=&\sum_{i=1}^M \frac{1}{2} \left( \hat{a}_i^\dagger \hat{b}_i + \hat{b}_i^\dagger \hat{a}_i \right),\\
\hat{S}_y&=&\sum_{i=1}^M \frac{1}{2i} \left( \hat{a}_i^\dagger \hat{b}_i - \hat{b}_i^\dagger \hat{a}_i \right),\\
\hat{S}_z&=&\sum_{i=1}^M \frac{1}{2} \left( \hat{a}_i^\dagger \hat{a}_i - \hat{b}_i^\dagger \hat{b}_i \right)
\end{eqnarray}
As for individual spins, collective spin components obey cyclic commutation relations $\comm{\hat{S}_x}{\hat{S}_y}=i \hat{S}_z$.
Due to the fact that the Bose-Hubbard Hamiltonian commutes with $\hat{S}_z$ and the initial state is an eigenstate of $\hat{S}_x$ one can write explicitly the expression for minimal fluctuations
\begin{eqnarray}
	 \Delta \hat{S}^2_{\perp, {\rm min}} &=& \frac{1}{2} \left[ 2 \langle \Delta \hat{S}_z^2\rangle + A - \sqrt{A^2 + B^2} \right],\\
	A&=&\langle \Delta \hat{S}_y^2\rangle-\langle \Delta \hat{S}_z^2\rangle,\\
	B &=& 2 {\rm Re}\left[ \langle \hat{S}_y \hat{S}_z \rangle - \langle \hat{S}_y \rangle \langle \hat{S}_z \rangle \right].\label{eq:ss_B}
\end{eqnarray}

\section{Construction and validity of two-mode model}\label{sec:2MM}

In this section we introduce a two-mode model that can be used to obtain analytical predictions~\cite{Kajtoch_2018}, and investigate the dependence of the best squeezing time on the lattice potential height. 
To construct the two-mode model in the most general way, we introduce the ground state energy $E_0(\hat{N}_a, \hat{N}_b)$ of the system having $N_a$ atoms in the internal state $a$ and $N_b$ atoms in $b$, and make its Taylor expansion up to the second order around the average values $\bar{N}_a$ and $\bar{N}_b$ of the atom numbers, in the initial state at $t=0^+$. This leads to
\begin{align}\label{eq:TEofEnergy}
    E_0&(\hat{N}_a, \hat{N}_b) = \varE_0 + \partial_a E_0 \, (\hat{N}_a - \bar{N}_a) +  \partial_b E_0 \, (\hat{N}_b - \bar{N}_b) + \nonumber \\
      &+ \frac{1}{2} \partial^2_{aa} E_0 \, (\hat{N}_a - \bar{N}_a)^2 
    + \frac{1}{2} \partial^2_{bb} E_0 \, (\hat{N}_b - \bar{N}_b)^2 + \nonumber \\
    &+\frac{1}{2} (\partial^2_{ab} E_0 + \partial^2_{ba} E_0) \, (\hat{N}_a - \bar{N}_a)(\hat{N}_b - \bar{N}_b) + \cdots ,
\end{align}
where $\varE_0\equiv E_0(\bar{N}_a, \bar{N}_b) $, $\partial_\sigma E_0 \equiv \frac{\partial E_0}{\partial N_\sigma}|_{\bar{N}_a, \bar{N}_b}$ and $\partial^2_{\sigma \sigma'} E_0 \equiv \frac{\partial^2 E_0}{\partial N_\sigma \partial N_{\sigma'}}|_{\bar{N}_a, \bar{N}_b}$, with $\sigma, \sigma'=a,b$. Notice, that the first derivative of the ground state energy is the zero temperature chemical potential $\mu_\sigma = \partial_\sigma E_0$ of the $\sigma $ component. By introducing the collective spin components, the ground state energy expanded to the second order (\ref{eq:TEofEnergy}) becomes \cite{Li2009}
\begin{equation}\label{eq:effectiveOATham}
    E_0(\hat{N}_a, \hat{N}_b) = f_N + \hbar v_N \hat{S}_z + \hbar \chi \hat{S}_z^2
\end{equation}
In the above equation, $f_N$ is a function of the total number of particles, the linear term in $\hat{S}_z$ describes spin precession around the $z$ axis with the velocity $v_N$, and finally 
\begin{equation}
    \chi = \frac{1}{2\hbar}\left( \partial_a \mu_a + \partial _b \mu_b - \partial _a \mu_b - \partial_b \mu_a \right).
\end{equation}
In the symmetric case that we consider, with $U_{aa}=U_{bb}$, one has $v_N=0$ (no precessions). 
Note that the approximation of the energy by its second order expansion is more and more accurate as $N$ is large, as the relative width of the distributions of $N_a$ and $N_b$ decrease as $1/\sqrt{N}$.

The approximated energy (\ref{eq:effectiveOATham}) is the basis of the two-mode model we consider here.
It has the form of the one-axis twisting (OAT) model introduced by Kitagawa and Ueda~\cite{PhysRevA.47.5138} 
and it catches the main features of the squeezing dynamics. In particular we expect that the maximum level of squeezing generated in our scheme is the same as the one generated by equivalent the OAT model (\ref{eq:effectiveOATham}), in both the static and the dynamic scheme that we consider. The time scale, or the best squeezing time, depends on the system Hamiltonian trough the parameter~$\chi$. Here below, we give an approximate expression~of~$\chi$ in the static case, assuming that all the atoms are in the zero momentum mode that is the condensate mode in the noninteracting case.
To this aim, we write the Bose-Hubbard Hamiltonian~(\ref{eq:BHM}) in the momentum representation, $\hat{a}_i=\frac{1}{\sqrt{M}}\sum_q e^{-i q x_i} \hat{a}_q$ and $\hat{b}_i=\frac{1}{\sqrt{M}}\sum_q e^{-i q x_i} \hat{b}_q$, and we keep only the zero momentum terms~\cite{PhysRevA.98.023621}. We obtain~:
\begin{align}\label{eq:energySF}
    \hat{\mathcal{H}}_{{\rm BH},\, 0}(\hat{N}_{0, a}, \hat{N}_{0,b}) &= 
    \epsilon_{0,a}\hat{N}_{0, a} + \epsilon_{0,b} \hat{N}_{0, b} \nonumber \\
    &+ \frac{U_{aa}}{2M} \hat{N}_{0,a}^2+ \frac{U_{bb}}{2M} \hat{N}_{0,b}^2 + \frac{U_{ab}}{M} \hat{N}_{0,a} \hat{N}_{0,b},
\end{align}
where $\hat{N}_{0,a}=\hat{a}^\dagger_{q=0} \hat{a}_{q=0}$, $\hat{N}_{0,b}=\hat{b}^\dagger_{q=0} \hat{b}_{q=0}$, $\epsilon_{0,a}=-2J-U_{aa}/(2M)$ and $\epsilon_{0,b}=-2J-U_{bb}/(2M)$.
The chemical potentials are 
$\mu_a|_{\bar{N}_{0,a},\bar{N}_{0,b}}=\epsilon_{0,a}+\frac{U_{aa}}{M}\bar{N}_{0,a}+\frac{U_{ab}}{M}\bar{N}_{0,b}$,
$\mu_b|_{\bar{N}_{0,a},\bar{N}_{0,b}}=\epsilon_{0,b}+\frac{U_{bb}}{M}\bar{N}_{0,a}+\frac{U_{ab}}{M}\bar{N}_{0,a}$ where all expressions are derived assuming that all atoms are in the zero momentum mode.
One then obtains
\begin{equation}\label{eq:chiq0}
    \chi_{q=0} = \frac{1}{2 M \hbar} (U_{aa} + U_{bb} - 2 U_{ab}), 
\end{equation}
Using (\ref{eq:chiq0}), the parameter $\chi_{q=0}$ can be calculated numerically using the exact form of the Wannier function to determine the interaction parameters $U_{\sigma\sigma'}$. Moreover, analytical expressions for $\chi_{q=0}$ can be obtained in the limiting cases of a very shallow or a very deep lattice. The first limit is that of a homogeneous system when $V_0\to 0$. The values of the interaction parameters are then $U^{\rm homo}_\sigma=g_\sigma N/L_x$, where $L_x=Md$ is the system size along $x$, leading to 
\begin{equation}
\hbar \chi^{\rm homo}_{q=0}= \frac{4}{M \pi} \frac{\Delta a}{d} E_R
\end{equation}
where $\Delta a = a_a + a_b -2 a_{ab}$. The second one is that of localized Wannier functions that can be approximated by Gaussians, see Appendix~\ref{app:JandU}. In this case one obtains 
\begin{equation}
\hbar \chi_{q=0}^{\rm Gauss}=\sqrt{\frac{32}{\pi}}\frac{\Delta a}{d}\left(\frac{V_0}{ E_R}\right)^{1/4} E_R
\end{equation}
The zero momentum part of the Bose-Hubbard Hamiltonian~(\ref{eq:energySF}) can be viewed as bimodal: the atoms in $q=0$ of the $a$ state and those in $q=0$ of the $b$ state. 
The evolution of such bimodal subsystem can be solved analytically in the Fock state basis:
\begin{equation}
 | \psi_0(t) \rangle =\sum\limits_{N_{0, a} = 0}^{N} c_{N_{0,a}} e^{-i\frac{\chi_{q=0} t}{4 \hbar} (N-2N_{0,a})^2 } |N_{0,a}, N-N_{0,a} \rangle,
 \label{eq:2m_time_evolution}
\end{equation} 
taking the spin coherent state as the initial state that determines the values of $c_{N_{0,a}}=2^{-N/2} \sqrt{\binom{N}{N_{0,a}}}$. In (\ref{eq:2m_time_evolution}) we omitted the constant global phase factor.
Consistently, for the two-mode model, the spin squeezing parameter (\ref{eq:spin_squeezing}) is considered for the zero momentum mode, replacing the collective spin operators $\hat{S}_{x}\, \hat{S}_{y}$ and $\hat{S}_{z}$ by their zero momentum counterparts 
$ \hat{S}_{0,x}=( \hat{a}_{q=0}^\dagger \hat{b}_{q=0} + \hat{b}_{q=0}^\dagger \hat{a}_{q=0} ) /2$, 
$ \hat{S}_{0, y}=( \hat{a}_{q=0}^\dagger \hat{b}_{q=0} - \hat{b}_{q=0}^\dagger \hat{a}_{q=0} ) /2i$ and $\hat{S}_{0,z}=(\hat{N}_{0,a} - \hat{N}_{0,b})/2$. The evolution of such spin squeezing parameter for zero-momentum-mode is solvable analytically~\cite{PhysRevA.47.5138,Alice.Sinatra.86}. We will refer to it as the two-mode model (2MM) while comparing it to the exact numerical results. We expect agreement between the exact calculation and the 2MM as long as $\langle \hat{N}_{0,a}\rangle+\langle \hat{N}_{0,b} \rangle \approx N$ in the initial state before $\pi/2$ pulse and as long as the system stays in the superfluid regime.

\begin{figure}[bt!]
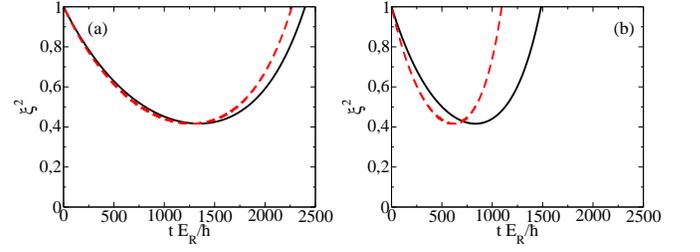

\centering
\includegraphics[width=.492\linewidth]{Fig2a.eps}
\includegraphics[width=.492\linewidth]{Fig2b.eps}
\caption{Squeezing parameter (\ref{eq:spin_squeezing}) versus time for $N=M=6$. (a) $V_0=0.4E_R$ and $f_c>0.99$. (b) $V_0=6E_R$ and $f_c=0.97$. The black solid lines show the exact numerical results while the red dashed lines the prediction of 2MM with $\chi_{q=0}$ from (\ref{eq:chiq0}). The agreement between the 2MM and exact results degrades for larger $V_0$.}
\label{fig:fig1}
\end{figure}

\begin{figure}[]
\centering
\begin{picture}(0,170)
\put(-117,00){\includegraphics[width=.95\linewidth]{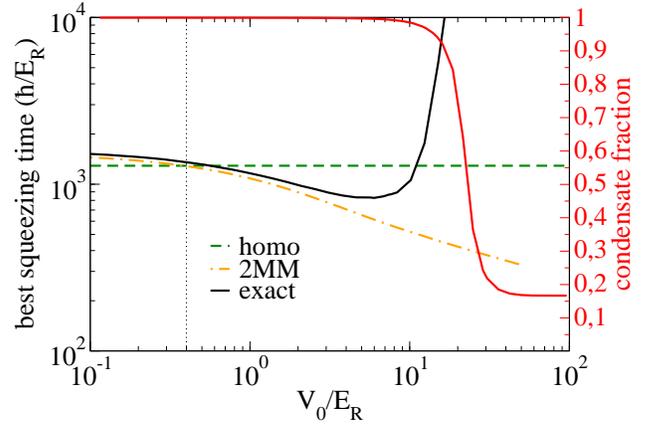}}
\end{picture}
\caption{Best squeezing time versus the static value of $V_0/E_R$ extracted from the exact numerical simulations (black solid line), from the 2MM (orange dot-dashed line) and for a homogeneous system (green dashed line). The corresponding condensed fraction (\ref{eq:condensate_fraction}) in state $a$ before the $\pi/2$ pulse (red solid line) is also shown, with the scale on the right axis. The vertical black dotted line marks $V_0/E_R=0.4$ where the presence of a lattice starts to accelerate the squeezing dynamics.}
\label{fig:fig2}
\end{figure}

In Fig.~\ref{fig:fig1} we show an example of the evolution of the spin squeezing parameter (\ref{eq:spin_squeezing}) for two different static values of the lattice height $V_0/E_R=0.4$ and $V_0/E_R=6$, both for $N=6$. 
The maximum level of squeezing achieved during the evolution is the same for the two values of $V_0/E_R$ while the time scales are different, as expected. The agreement between the 2MM and the exact results degrades when $V_0$ increases,
because the approximation of having all the atoms in the zero momentum mode, that we used to calculate $\chi_{q=0}$, becomes unjustified. 
To be more specific on this point, in Fig.~\ref{fig:fig2} we collect the best squeezing times calculated exactly (black solid line) and compare them to the 2MM results (orange dot-dashed line) with $\chi_{q=0}$ calculated from (\ref{eq:chiq0}). The condensed fraction defined as the average occupation of the $q=0$ momentum mode:
\begin{equation}\label{eq:condensate_fraction}
    f^{\sigma}_c=\frac{1}{N} \langle \hat{\sigma}^\dagger_{q=0} \hat{\sigma}_{q=0}\rangle=\frac{1}{NM} \sum_{i,j}\langle \hat{\sigma}^\dagger_{i} \hat{\sigma}_{j}\rangle ,
\end{equation}
where $\sigma=a,b$
is shown on the same figure. 
The following observations can be made: (i) the 2MM works well as long as the condensed fraction is close to one, and (ii) the best squeezing time is smaller than for an homogeneous system with the same number of atoms in a specific range of $V_0$. The squeezing acceleration by the optical lattice potential in the range $0.4E_R<V_0<7E_R$, is simply explained in the following paragraph.

Lets us first consider the condition $t_{\rm best} < t^{homo}_{\rm best}$.  It implies $\chi^{homo} < \chi$ and hence $\frac{\lambda}{2} \int |w|^4 dx>1$. By approximating the Wannier function $w$ by a Gaussian one obtains the lower bound for the lattice height $V_0/E_R>(\frac{2}{\pi})^2\approx 0.4$. This number is in a good agreement with what we observe in the numerical calculation, see vertical dotted line in Fig.~\ref{fig:fig2}. The upper bound for $V_0$ expresses the fact that the lattice height should be low enough for the system to remain in the superfluid phase with a condensed fraction close to one. To obtain an estimate of that for small atom numbers, we can use the explicit analytical expression of the condensed fraction for $N=2$, see Appendix~\ref{app:Neq2}. It is $f^a_c= \frac{64J^2 +16J \Omega_{aa+} +\Omega_{aa+}^2}{2 \left(64J^2+ \Omega_{aa+}^2\right)}$ with $\Omega_{aa+}=U_{aa}+ \sqrt{64J^2 +U_{aa}^2}$. 
It is obvious that $f^a_c \to 1$ when $U_{aa}/J \to 0$. In order to estimate the validity range of the 2MM we can fix the condensed fraction to a value close to one. By taking for example $f^a_c\approx 0.99$, the threshold point above which the 2MM fails is then $(U_{aa}/J)_{th}\approx 1.6$ corresponding to $(V_0/E_R)_{th}\approx 7$, which gives a fair estimate of what we see in the exact numerical calculations in Fig.~\ref{fig:fig2}.

\section{Spin-squeezed and GHZ states stored in Mott phase}\label{sec:structure}

\begin{figure}[bt!]
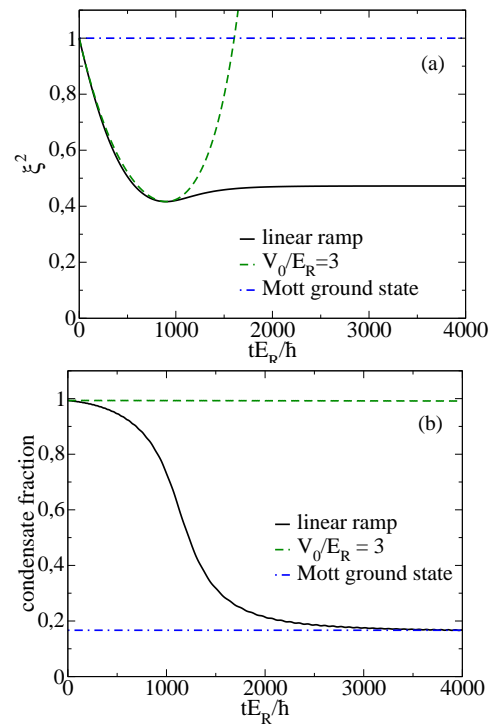

\centering
\includegraphics[width=.73\linewidth]{Fig4a.eps}
\includegraphics[width=.72\linewidth]{Fig4b.eps}
\caption{(a) Squeezing parameter $\xi^2$ and (b) condensed fraction $f_c^a$, for $N=6$, are shown (i) across the lattice ramp (\ref{eq:ramp1}) with $\tau E_R/\hbar=4255$ (black solid lines), and (ii) in a static situation for $V_0/E_R=3$ (green dashed lines). The same quantities in the $\pi/2$ ``Mott-insulator phase-state" (\ref{eq:MI_pi2}), for which $\xi^2_{\rm MI \, \pi/2}=1$ and $f^a_c{}_{\rm MI \, \pi/2}=1/N$, are shown for comparison (blue dot-dashed lines).}
\label{fig:fig3}
\end{figure}

In order to store the best squeezing in the Mott-insulator phase, we linearly ramp the lattice potential depth according to (\ref{eq:ramp1}). We choose the initial value of the lattice depth $V_0=3E_R$ as it corresponds to a condensed fraction close to unity, and it generates the squeezing faster than for a homogeneous system. An example of spin squeezing parameter and condensed fraction evolution\footnote{The values of the tunneling and interaction terms were calculated exactly according to (\ref{eq:tunneling}) and to (\ref{eq:u_definition}). They are shown in Fig.~\ref{fig:fig30}} is shown in Fig.~\ref{fig:fig3}.
The squeezing parameter stops evolving around its minimum because the ramp is adjusted so that the system undergoes the Mott transition at the best squeezing time. In the Mott-insulator phase with one atom per site, atoms are isolated around lattice sites and stop to interact. As we will show it in this section, not only a squeezed state can be stored in the Mott phase. By varying the ramp time $\tau$ one can store other states dynamically produced by the interactions, including macroscopic superpositions of phase states.

To gain understanding about the structure of state stored in Mott phase, lets us first consider the simplest case of $N=2$ and then generalize the arguments for an arbitrary number of atoms. 

\subsection{Case $N=2$}

\begin{figure}[bt!]
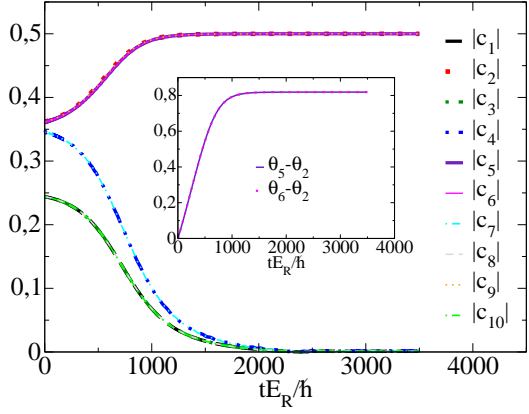

\centering
\begin{picture}(0,155)
\put(-110,0)
{\includegraphics[width=0.8\linewidth]{Fig5a.eps}}
\put(-55,50)
{\includegraphics[width=0.4\linewidth]{Fig5b.eps}}
\end{picture}
\caption{Time evolution of the coefficients in the state (\ref{eq:generalstate}), for $N=2$ and $\tau E_R/\hbar=4255$. 
The moduli $|c_w(t)|$ are in the main figure while the inset shows the relative phases of the $c_w(t)$ that are nonzero in the long time limit. Here we introduced the notation $c_w=|c_w|e^{i \theta_w}$.}
\label{fig:fig40cw}
\end{figure}

In general, the state of the system can be considered in the Fock state basis as we did numerically. For $N=2$ it reads
\begin{equation}\label{eq:generalstate}
    |\Psi(t) \rangle = \sum_{w=1}^{10} c_w(t) |w\rangle,
\end{equation}
where
\begin{align}\label{eq:states2}
    \ket{1}&=\ket{20,00}, \, 
    \ket{2}=\ket{11,00}, \,
    \ket{3}=\ket{02,00},\nonumber \\
    \ket{4}&=\ket{10,10},\,
    \ket{5}=\ket{10,01},\,
    \ket{6}=\ket{01,10},\,
    \ket{7}=\ket{01,01},\nonumber \\
    \ket{8}&=\ket{00,20},\,
    \ket{9}=\ket{00,11},\,
    \ket{10}=\ket{00,02}.
\end{align}
The coefficients of the decomposition in the Fock basis satisfy $c_6=c_5$, $c_7=c_4$, $c_8=c_3$, $c_9=c_2$, $c_{10}=c_1$,
$c_3=c_1$ and $c_8=c_{10}$, as a consequence of the symmetry $a \leftrightarrow b$ with respect to the exchange of the internal states and of the symmetry with respect to exchange of the lattice sites.
The evolution of the coefficients can be found analytically when $U_{aa}$ and $J$ are time-independent as discussed in Appendix~\ref{app:Neq2}. In the time-dependent case, the evolution of the $c_w$ can be treated numerically and the result is shown in Fig.\ref{fig:fig40cw}.

\begin{figure}[t!]
\centering
\includegraphics[width=1.\linewidth]{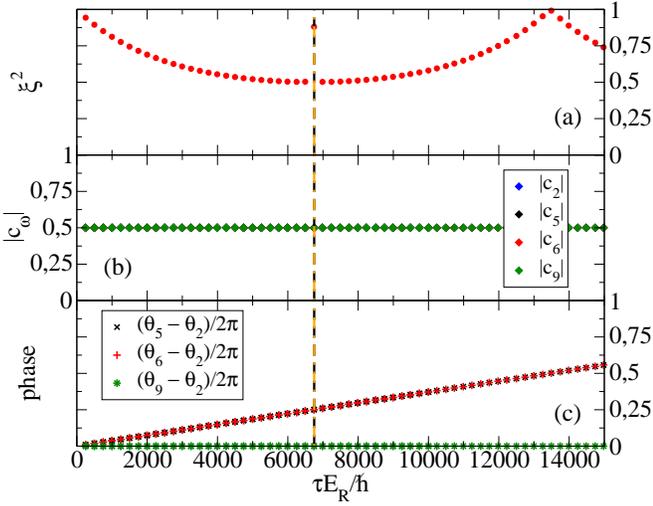}
\caption{
$N=2$. (a) Squeezing parameter versus ramp time $\tau$, (b) corresponding amplitudes and (c) relative phases divided by $2\pi$ of nonzero coefficients of the state decomposition in the Fock basis, in the long time limit after stabilization in the Mott phase. The value of the squeezing parameter stored depends on the ramp time showing that an appropriate choice of $\tau$ is crucial to efficiency of the protocol. The dashed black vertical line marks the best squeezing. The only nonzero coefficients are $c_2, \, c_5,\, c_6$ and $c_9$
and the nonzero relative phases~: $\theta_5- \theta_2$ and $\theta_6- \theta_2$ are equal, demonstrating validity of the state (\ref{eq:SSM}). This relative phase, marked by $\phi$ in (\ref{eq:SSM}), linearly depends on $\tau$. The vertical dot-dashed orange line marks the ramp time $\tau$ corresponding to $\phi=\pi/2$ for storing the $N=2$ even superposition state (\ref{eq:SSMphi1phi2}).}
\label{fig:fig40}
\end{figure}

It turns out that the state stored in the Mott-phase has an interesting structure. The coefficients $c_w$ are nonzero only for Fock states that have one atom per lattice site. 
All the nonzero coefficients have identical absolute values and the only difference among them is in the phase factor. Referring to the states (\ref{eq:states2}), we find numerically that the state stored in the Mott phase at the end of the ramp for $N=2$ has the form
\begin{equation}\label{eq:SSM}
     |\Psi\rangle_{{\rm MI}, N=2} =\frac{1}{2} \left(\ket{2} + e^{i\phi}\ket{5} + e^{i\phi}\ket{6} +\ket{9} \right)
\end{equation}
up to a constant global phase factor. The only nonzero coefficients in the long time limit are $c_2, \, c_5, \, c_6$ and $c_9$. The very same Fock states appear in what we shall call the ideal $\phi=0$ Mott-insulator phase-state,
\begin{equation}
|\phi=0 \rangle_{\rm MI} =
\frac{\left( a_1^\dagger+b_1^\dagger \right)}{\sqrt{2}}
\frac{\left( a_2^\dagger+b_2^\dagger \right)}{\sqrt{2}}
|0\rangle = \frac{|2\rangle + |9\rangle +|5\rangle +|6\rangle}{2}
\label{eq:MI_pi2}
\end{equation}
obtained from a single component Mott state, by putting each atom in the coherent-superposition 
$(|a\rangle+|b\rangle)/\sqrt{2}$.
However, unlike in the $\phi=0$ Mott-insulator phase-state, a nonzero relative phase appears between individual Fock states. The value of the relative phase $\phi$ depends on the state that is stored in the Mott phase.
The above introduced structure (\ref{eq:SSM}) in term of Fock states, is illustrated in Fig.~\ref{fig:fig40} for different values of the ramp time, corresponding to different quantum states stored in the Mott-insulator phase. 

The general form (\ref{eq:SSM}) of the stored state for $N=2$ can be expressed in terms of two-site phase-states,
\begin{equation}
|\phi_1,\phi_2\rangle =
\frac{\left( a_1^\dagger+e^{i\phi_1}b_1^\dagger \right)}{\sqrt{2}}
\frac{\left( a_2^\dagger+e^{i\phi_2}b_2^\dagger \right)}{\sqrt{2}}
|0\rangle.
\label{eq:phi1phi2}
\end{equation}
Indeed one has
\begin{equation}\label{eq:SSMphi1phi2}
    |\Psi\rangle_{{\rm MI}, N=2} =e^{i\phi/2} \left({\rm cos}\frac{\phi}{2}\ket{0,0} 
    + i {\rm sin}\frac{\phi}{2}\ket{\pi,\pi} \right).
\end{equation}
It is true for any value of the phase $\phi$. In this sense, the state stored in the Mott phase is a superposition of two different Mott-insulator phase-states with different phases, and it becomes an even superposition when $\phi=\pi/2$.

\begin{figure}[t!]
\centering
\includegraphics[width=1.\linewidth]{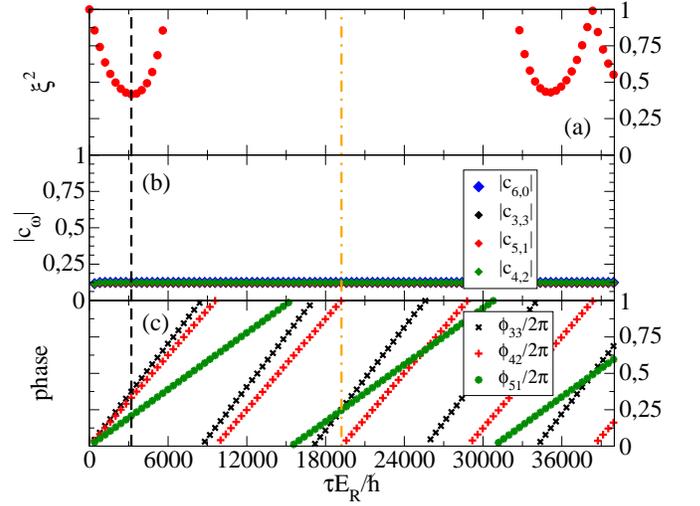}
\caption{
$N=6$. (a) Squeezing parameter versus ramp time $\tau$, (b) corresponding amplitudes and (c) relative phases divided by $2\pi$ of nonzero coefficients of the state decomposition in the Fock basis, in the long time limit after stabilization in the Mott phase. There are 64 nonzero coefficients $c_w$ whose absolute values are close to $2^{-3}$. The three different relative phases between nonzero coefficients, $\phi_{33}$, $\phi_{42}$, $\phi_{51}$ in (\ref{eq:SSMN6}), linearly depend on $\tau$ with individual slopes. The ramp times $\tau$ for storing the best squeezed and the GHZ state are marked by the vertical black dashed and orange dot-dashed lines, respectively.}
\label{fig:fig400}
\end{figure}

\subsection{Case $N=6$}
The natural question arises how the discussed above structure changes when the number of atoms increases. One might expect that the state stored in the Mott phase is a superposition of $2^N$ Fock states with single occupation per lattice site and $N/2$ different phase factors. We have checked numerically that this is the case for $N=4$ and $N=6$. In the latter case one has
\begin{align}
    \ket{\Psi}_{{\rm MI},N=6}=&\frac{1}{2^3}\left[ 
    \mathcal{S}_{6,0} + \mathcal{S}_{0,6} 
    +e^{i\phi_{51}}
    \left( \mathcal{S}_{5,1} + \mathcal{S}_{1,5} \right) \right.\nonumber \\
    +&e^{i\phi_{42}}
    \left( \mathcal{S}_{4,2} + \mathcal{S}_{2,4} \right)
    +e^{i\phi_{33}}\mathcal{S}_{3,3} \left.  \right],
    \label{eq:SSMN6}
\end{align}
were we introduced notation
\begin{equation}
    \mathcal{S}_{N_a,N_b} = 
    \sum_{\{ n_{ia}\}^\prime \atop \sum_i n_{ia}=N_a} \,
    \sum_{\{ n_{ib}\}^\prime \atop \sum_i n_{ib}=N_b}
    \ket{\{ n_{ia}\}, \{ n_{ib}\}},
\end{equation}
in which summations run over all configurations $\{ n_{ia}\}$ and $\{ n_{ib}\}$ of occupation numbers on the lattice in the state $a\, (b)$, under the conditions of a fixed number of atoms in state $a$ and $b$, and the restriction of single occupation per lattice site indicated by the prime. The state $\ket{ \{ n_{ia}\}, \{ n_{ib}\}}$ denotes the multi-site Fock state.
For example $\mathcal{S}_{6,0}=\ket{111111,000000}$. 
The structure (\ref{eq:SSMN6}) of the state for $N=6$, is illustrated in Fig.~\ref{fig:fig400} for different values of the ramp times, corresponding to different quantum states stored in the Mott-insulator phase. 

As we did it previously, we can express the stored state in terms of Mott-insulator multi-site phase-states. In general, in the case of six atoms and six lattice sites, the state (\ref{eq:SSMN6}) can be expressed in terms of six-site phase-states
\begin{align}
    | \phi_1, \phi_2, \phi_3, \phi_4, \phi_5, \phi_6 \rangle 
    &= \prod_{j=1}^6 \frac{\hat{a}_j^\dagger + e^{i \phi_j}\hat{b}_j^\dagger}{\sqrt{2}} |0\rangle \nonumber \\
    &=\frac{1}{2^3}\sum_{N_a=0}^6 \tilde{\mathcal{S}}_{N_a, N-N_a},
    \label{eq:N6phasestate}
\end{align}
with
\begin{equation}
    \tilde{\mathcal{S}}_{N_a, N_b} = 
    \sum_{\{ n_{ia}\}^\prime \atop \sum_i n_{ia}=N_a} \,
    \sum_{\{ n_{ib}\}^\prime \atop \sum_i n_{ib}=N_b}
    e^{i\sum_i \phi_i n_{bi}} \ket{\{ n_{ia}\}, \{ n_{ib}\}}.
\end{equation}
When considering a Mott-insulator phase-state $|\phi\rangle_{MI}$ (\ref{eq:N6phasestate}) where the phase is the same for all lattice sites, 
\begin{equation}
| \phi \rangle_{MI} \equiv | \phi, \phi, \phi, \phi, \phi, \phi \rangle
\end{equation}
we can replace in (\ref{eq:N6phasestate}) $\tilde{\mathcal{S}}_{N_a, N_b}$ by $e^{i\phi N_b} \mathcal{S}_{N_a, N_b}$.
The different terms in (\ref{eq:SSMN6}) can then be expressed with the help of Mott-insulator phase-states as
\begin{align}
\frac{3}{4}\left(\mathcal{S}_{60}+\mathcal{S}_{06} \right)
&= |0\rangle_{MI} + |\pi \rangle_{MI} \nonumber \\ &+  |\frac{\pi}{3} \rangle_{MI} + |\frac{2\pi}{3} \rangle_{MI} + |\frac{4\pi}{3} \rangle_{MI} + |\frac{5\pi}{3} \rangle_{MI} , \nonumber \\
 \frac{1}{2}\left(\mathcal{S}_{51}+\mathcal{S}_{15} \right)
&= |0\rangle_{MI} - |\pi \rangle_{MI}  \nonumber\\
& - i \left( |\frac{\pi}{2} \rangle_{MI} + |\frac{3\pi}{2} \rangle_{MI} \right), \nonumber \\
 \frac{3}{4}\left(\mathcal{S}_{42}+\mathcal{S}_{24} \right)
&= 2|0\rangle_{MI} + 2|\pi \rangle_{MI}  \nonumber\\
& - \left( |\frac{\pi}{3} \rangle_{MI} + |\frac{2\pi}{3} \rangle_{MI} + |\frac{4\pi}{3} \rangle_{MI} + |\frac{5\pi}{3} \rangle_{MI} \right), \nonumber \\
 \frac{1}{2} \mathcal{S}_{33}
&= |0\rangle_{MI} - |\pi \rangle_{MI}  \nonumber\\
&+ i \left( |\frac{\pi}{2} \rangle_{MI} + |\frac{3\pi}{2} \rangle_{MI} \right) \,.
\end{align}
In the special case when $e^{i\phi_{42}}=1$ and $\phi_{51}=\phi_{33}=\pi/2$, 
one obtains the $N=6$ Greenberger-Horne-Zeilinger state
\begin{equation}\label{eq:GHZN=6}
    \ket{\Psi}_{{\rm GHZ},N=6}=\frac{e^{i \pi/4}}{\sqrt{2}} 
    \left[ | 0 \rangle_{MI} + i | \pi \rangle_{MI} \right] .
\end{equation}
As shown in Fig.~\ref{fig:fig400}, this occurs when the ramp time $\tau$ equals half of period of the one-axis twisting dynamics, corresponding to freezing in the Mott state the correlations of a Schr\"odinger cat, macroscopic superposition of two phase-states.  
Note that (\ref{eq:GHZN=6}) is equivalent to the most common representation of the Greenberger-Horne-Zeilinger state in the number operator basis. To see this, one should change the basis by rotation of the state (\ref{eq:GHZN=6}) by $\pi/2$ around the $y$-axis.

\subsection{On-site and inter-site correlation functions in the Mott-squeezed state}

The stored Mott-squeezed state is neither a $\phi=0$ Mott-insulator phase-state nor a two-mode squeezed state. It is a multimode entangled state possessing some of the properties the two.
To have more insight into the character of the Mott-squeezed state, we analyze here the two correlation functions: $\langle \hat{a}^\dagger_i \hat{b}_j \rangle$ and 
$\langle \{\hat{S}_{y,i} , \hat{S}_{z, j} \} \rangle - 2\langle \hat{S}_{y,i} \rangle\langle \hat{S}_{z,j}\rangle $, 
where $\hat{S}_{y,i} = (\hat{a}^\dagger_i\hat{b}_i - \hat{b}^\dagger_i\hat{a}_i)/2i$ and 
$\hat{S}_{z,i} = (\hat{a}^\dagger_i\hat{a}_i - \hat{b}^\dagger_i\hat{b}_i)/2$.
The first one quantifies the correlation between the internal states $a$ and $b$ of the atoms, that is maximal for $i=j$ in a $\phi=0$ Mott-insulator phase-state. The second one characterizes spin-spin correlations between different atoms, that is a property of spin-squeezed states
as can be seen by the definitions (\ref{eq:spin_squeezing}) and (\ref{eq:ss_B}). The higher the value of the spin-spin correlation function the higher the level of squeezing. Note that $\langle \hat{S}_{y,i} \rangle=\langle \hat{S}_{z,i}\rangle =0$ both for the initial state and during evolution. 

\begin{figure}[bt!]
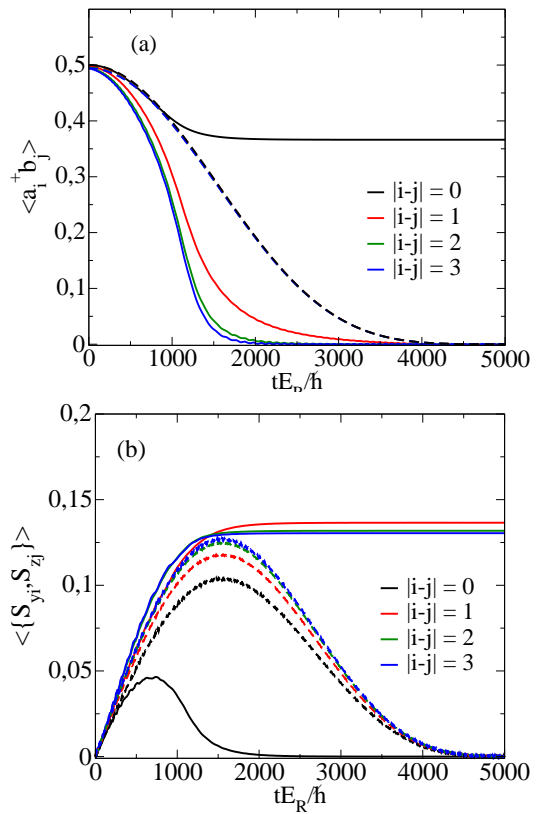

\centering
\includegraphics[width=.8\linewidth]{Fig8a.eps}
\includegraphics[width=.8\linewidth]{Fig8b.eps}
\caption{
(a) Internal states correlation functions $|\langle \hat{a}_i^\dagger \hat{b}_j \rangle |^2$ and (b) spin-spin correlation functions $\langle \{\hat{S}_{y,i} , \hat{S}_{z, j} \} \rangle$, for the linear ramp (solid lines). For comparison, the same correlation functions are also shown for the static situation corresponding to the initial value of $V_0/E_R=3$ (dashed lines). Correlation functions between different sites $i-j$ are marked by different colors as indicated in the legends.}
\label{fig:fig4}
\end{figure}

In Fig.~\ref{fig:fig4} we show both correlation functions for different distances $i-j$ between the sites, both in the case of a linear ramp of the lattice (solid lines) and in the case of a static lattice (dashed lines). In the case of a ramp, in
the long time limit when $J\to 0$, the $|\langle \hat{a}_i^\dagger \hat{b}_j \rangle |^2$ correlation function is nonzero only on-site;  the off-site correlation function $\langle \hat{a}^\dagger_i \hat{b}_j \rangle$ tends to zero independently of the value $i-j$ as long as $i\ne j$. On the other hand, the spin-spin correlation function is nonzero only between different sites $i\ne j$.

The origin of such a behavior comes from the structure of the Mott-squeezed state, and can be understood in detail in the case $N=2$. The correlation functions of interest are in this case
\begin{align}
    \langle \hat{a}^\dagger_1 \hat{b}_1\rangle&
    =2\sqrt{2}|c_1c_4| {\rm cos}(\theta_{1}-\theta_{4})
    +2|c_2c_5|{\rm cos}(\theta_{2}-\theta_{5}), \label{eq:abcorr11}\\
    \langle \hat{a}^\dagger_1 \hat{b}_2\rangle& 
    =2\sqrt{2}|c_1c_5| {\rm cos}(\theta_{1}-\theta_{5})
    +2|c_2c_4|{\rm cos}(\theta_{2}-\theta_{4}),  \label{eq:abcorr12}
\end{align}
and
\begin{align}
    \langle \{\hat{S}_{y,1} , \hat{S}_{z, 1} \} \rangle&
    =2\sqrt{2} |c_1c_4|{\rm sin}(\theta_4 - \theta_1) ,\label{eq:SScorr11}\\
    \langle \{\hat{S}_{y,1} , \hat{S}_{z, 2} \} \rangle& 
    =2|c_2 c_5| {\rm sin}(\theta_{5}-\theta_{2}),\label{eq:SScorr12}
\end{align}
where we used (\ref{eq:generalstate}), the $a \leftrightarrow b$ symmetry  
and the notation $c_w=|c_w|e^{i\theta_{w}}$.
Since with one atom per site, only the coefficients $c_2,c_5,c_6,c_9$ are nonzero,
the analytical expressions (\ref{eq:abcorr11})-(\ref{eq:abcorr12}) confirm that $\langle \hat{a}^\dagger_i \hat{b}_j\rangle$ is nonzero when $i=j$ and zero for $i\ne j$. One can check that the product of nonzero coefficients (those of Fock states with one atom per lattice site) always contribute to the on-site function $\langle \hat{a}^\dagger_i \hat{b}_i\rangle$, independently of the total atom number as long as unit filling is considered. 
Similarly one can see from (\ref{eq:SScorr11}) and (\ref{eq:SScorr12}) that the spin-spin correlation function has the opposite behavior, displaying a nonzero product of coefficients when $i\ne j$ only
\footnote{From the mathematical point of view, it is because the two nonzero terms in the anti-commutator in (\ref{eq:SScorr11}) and (\ref{eq:SScorr12}) compensate each other for $i=j$. In opposite, for the $j\ne j$ case they both contribute to the function as they have the same sign.}.

\section{Effect of nonzero temperature on the Mott-squeezed state}\label{sec:T}

Two types of excitations exist in the Mott phase of our two component system: gapped excitations corresponding to double occupations of a site, for which the energy scale is the on-site interaction energy $U_{\sigma \sigma}$ or $U_{\sigma \sigma'}$, and ``soft" excitations within the single occupation manifold that can be described by an effective spin model. 
In our protocol it is important to maintain the system in the single occupation manifold in order to keep the stored squeezing and entanglement constant in time. Within this manifold, it is not important to remain in the ground state because the spin-dependent interaction energy is strongly reduced, scaling as $J^2/U_{\sigma \sigma'}$~\cite{PhysRevLett.90.100401, Altman_2003, PhysRevA.79.053614, PhysRevB.80.245109}. 

We therefore concentrate on excitations out of the low-energy manifold toward states with double occupations. We show that thermal fluctuations cause the squeezing parameter to oscillates in the Mott phase. Its value nevertheless remains small in the low temperature limit.
Residual interaction  in the Mott phase due to double occupation of a site, directly influences our scheme because the squeezing dynamics is then imperfectly stopped.
To give a rough estimate, we expect that thermal effects become visible when the temperature is comparable with the excitation energy for double site occupation $k_BT \ge U_{ab} \simeq U_{bb}=U_{aa}$, where $k_B$ is the Boltzmann constant. 

\begin{figure}[]
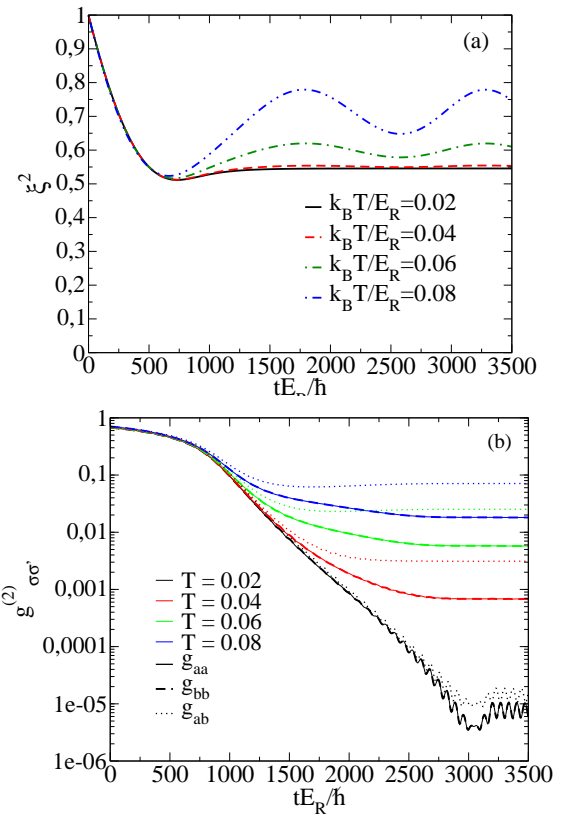

\centering
\includegraphics[width=0.8\linewidth]{Fig9a.eps}
\includegraphics[width=0.85\linewidth]{Fig9b.eps}
\caption{(a) Squeezing parameter versus time at nonzero temperature  and (b) second order correlation function.
Here $N=4$. 
The values of the initial condensed fractions are $f_c^a=0.997 \, (k_B T/E_R=0.02)$, 
$f_c^a=0.996 \, (k_B T/E_R=0.04)$, $f_c^a=0.989 \, (k_B T/E_R=0.06)$ and $f_c^a=0.973 \, (k_B T/E_R=0.08)$.
As long as the initial condensed fraction is large, and the initial temperature is week with respect to the value of $U_{ab}$ and $U_{aa}$ (as discussed in the text), the effect of the temperature is very weak.
}
\label{fig:fig10}
\end{figure}

According to Fig.\ref{fig:fig30} and Fig.\ref{fig:fig3}, we see that $U_{aa}$ varies between 
$0.065E_R$ and $0.08E_R$ in the critical region from $tE_R/\hbar=1000$ and $tE_R/\hbar=2000$, after which the thermally induced double occupations are frozen in the deep Mott phase. We then expect notable thermal effects for temperatures of this order. The finite temperature simulation results are shown in 
Fig.~\ref{fig:fig10}(a) for $N=4$. We observe that nonzero temperature not only increases the value of the best squeezing stored in the Mott phase as it was already understood for bimodal condensate~\cite{PhysRevLett.107.060404}; in addition, an evolution of the squeezing parameter stored in the Mott phase takes place in the form of regular oscillations. 
We expect the amplitude of oscillations of $\xi^2(t)$ in the Mott phase to be proportional to the  double occupation probability, quantified by the second order correlation function $g^{(2)}_{\sigma, \sigma'}=\langle \hat{\sigma}_i^\dagger \hat{\sigma'}_i^\dagger \hat{\sigma}_i \hat{\sigma'}_i \rangle/\sqrt{\langle \hat{\sigma}_i^\dagger \hat{\sigma}_i \rangle \langle \hat{\sigma'}_i^\dagger \hat{\sigma'}_i \rangle}$ with $\sigma,\sigma'=a,b$, illustrated in Fig.\ref{fig:fig10}(b). Indeed, while the temperature changes from
$0.06E_R$ to $0.08E_R$, both the stationary value of the second order correlation function and the oscillation amplitude of the squeezing parameters increase by approximately the same factor. On the other hand, the period of the oscillations in $\xi^2(t)$ in the Mott phase is determined by the periodicity of the dynamics for two atoms on the same site, similarly to \cite{Mandel03}, corresponding here to $\varphi=(U_{aa}-U_{ab}) t/\hbar=2\pi$.
Using the value of $U_{aa}$ at the end of ramp, for $V_0/E_R=40$ one finds a value of the period of approximately $1369\hbar/E_R$
~\footnote{At the end of the ramp, $U_{bb}=U_{aa}/E_R\approx 0.09$, and $U_{ab}=0.95U_{aa}$. One then finds $\Delta t = 2\pi \hbar /(U_{aa}-U_{ab}) = 1369.22 E_R/\hbar$.} in agreement to what observed in Fig.~\ref{fig:fig10}. 

Our results demonstrate the possibility to maintain spin-squeezing in the Mott phase in our small system even at finite temperature.

\section{Summary}\label{sec:conclusion}

We investigated the dynamical generation and storage of spin-squeezed and Greenberger-Horne-Zeilinger states in the Mott phase of system of a cold atoms in an optical lattice. This is done by raising an optical the lattice potential in an interacting bimodal Bose-Einstein condensate that is initially in the superfluid phase in the lattice. 

We introduced and explained a two-mode one-axis twisting model which allows analytical estimates, and used it to show that the presence of the optical lattice can accelerate the entanglement dynamics as compared to an homogeneous system. The careful analysis of the stored entangled state in the Mott phase revealed its structure:  it is composed by the same Fock states that are present in an ideal $\phi=0$ Mott-insulator phase-state, but with additional relative phase factors depending on the state that is stored. We show that by a proper choice of the lattice ramp time, that would correspond to formation of a Schr\"odiger cat in a two-mode one-axis twisting model, it is possible to store in the lattice a Greenberger-Horne-Zeilinger state. The structure of the Mott-squeezed states was used to explain the behavior of several correlation functions, on-site and between different sites, that we have calculated numerically.
The first-order internal state correlation functions $\langle \hat{a}^\dagger_i \hat{b}_j \rangle$  are nonzero only when $i=j$ demonstrating one-body coherence within each atom. The spin-spin correlation functions $\langle \{\hat{S}_{y,i} , \hat{S}_{z, j} \} \rangle$, that characterize spin-squeezing are nonzero only when $i\ne j$ as a consequence of unit filling. Finally, we discussed the effect on the squeezing of nonzero temperature that allows particle-hole excitations. We show that it is still possible to store a squeezed state in the Mott phase, although oscillations of the squeezing parameter appear at finite temperature due to the nonzero probability of double occupation of a single site.

From the experimental point of view, the system with a few atoms is very interesting because of the large control one has on the system, both for the preparation and for the measurement.  A step-by-step scaling up of the system size, by increasing the number of lattice sites or the spatial dimension, would allow to explore the boundary between the microscopic and the macroscopic world.

\section*{ACKNOWLEDGMENTS}
We thank D. Kajtoch for very initial contribution to the project. This work was supported by the Polish National Science Center Grants DEC-2015/18/E/ST2/00760 (MK) and under QuantERA programme, Grant No. UMO-2019/32/Z/ST2/00016 (EW). This project has received funding from the QuantERA Programme under the acronym MAQS.
A.S. acknowledges support from the Horizon 2020 European Project macQsimal. 
M.P. was supported by the Foundation for Polish Science through the IRA Programme co-financed by EU within SG OP.

\appendix

\section{Bose-Hubbard model}\label{app:BHmodel}

The motion of a particle in a periodic potential formed by an optical lattice is conveniently described by the band theory, and the system Hamiltonian (\ref{eq:Hamiltonian}) can be considered in the basis of Bloch functions $\psi_{l,q}(x, t)$~\cite{RevModPhys.80.885}. These are eigenfunctions of the single-particle Hamiltonian (\ref{eq:single_particle_hamiltonian}) and posses required transnational properties. The Bloch functions can be constructed numerically in the plane wave basis as described in Appendix~\ref{ap:bloch}. As a next step, the Hamiltonian is rewritten in the basis of Wannier functions $w(x-x_i,t)$ localized around lattice sites, where $x_i$ denotes position of the $i$-th site in the lowest energy band. The Wannier functions are conveniently constructed from the Bloch states $\psi_{l=1,q}(x,t)$ in the following way
\begin{equation}\label{eq:wannier_bloch}
w(x-x_i,t) = \left( \frac{d}{2 \pi} \right) ^{1/2} \int\limits_{q \in BZ} {\rm d}q \, e^{-i q x_i } \psi_{l=1,q}(x,t).
\end{equation}
Summation in Eq.~\eqref{eq:wannier_bloch} extends over wave vectors belonging to the 1st Brillouin zone (BZ), $-k< q \leq k$. The two-component Hamiltonian is then
\begin{align}\label{eq:ham}
\hat{\mathcal{H}} &= -\sum\limits_{i, j} J(i-j) \left(\hat{a}_{i}^{\dagger}\hat{a}_{j} + \hat{b}_{i}^{\dagger}\hat{b}_{j}\right) \nonumber\\
+ & \frac{1}{2}\sum\limits_{i,j,k,l} U^{aa}_{i,j,k,l} \hat{a}^\dagger_i \hat{a}^\dagger_j \hat{a}_k \hat{a}_l 
+ \frac{1}{2}\sum\limits_{i,j,k,l} U^{bb}_{i,j,k,l} \hat{b}^\dagger_i \hat{b}^\dagger_j \hat{b}_k \hat{b}_l \nonumber\\
 + & \sum \limits_{i,j,k,l} U^{ab}_{i,j,k,l} \hat{a}^\dagger_i \hat{b}^\dagger_j \hat{a}_k \hat{b}_l ,
\end{align}
where $\hat{a}_i^{\dagger} (\hat{b}_i^{\dagger})$ creates a particle in the single-particle Wannier state $w(x-x_i,t)$ of the lowest energy band ($l=1$) localized on the $i$-th site, in the internal state $a \, ( b )$. 
The Bose-Hubbard model considers only states in the lowest energy band, which is justified as long as the excitations energies to the higher bands are much larger than energies involved in the system dynamics. In general, if $V_{0}(t)$ is varied in time, the Wannier states, and hence the hoping and interaction parameters depend on time
\begin{subequations}\label{eq:hoping_inter} 
 \begin{align}
 J(i-j) & = - \frac{d}{2\pi} \int\limits_{q \in BZ} {\rm d} q \, E_q e^{-i (i-j)d q}\, \label{eq:tunneling}\\
  U^{\sigma \sigma'}_{i,j,k,l} & = g_{\sigma \sigma'} \int {\rm d}x\, w(i)w(j)w(k)w(l),
  \label{eq:u_definition}
  \end{align}
\end{subequations}
were we introduced $w(i)=w(x-x_i)$. 
In Appendix \ref{app:JandU} we discuss how the different tunneling (\ref{eq:tunneling}) and interaction (\ref{eq:u_definition}) terms depend on the lattice height $V_0$. 

In the tight-binding limit when the lattice height is larger than the recoil energy $E_R=\hbar^2k^2/(2m)$ and the Wannier functions are well localized around each lattice site, the tunneling and interactions terms fall-off rapidly with the distance $|x_i - x_j|$. By keeping  only the leading terms, on obtains the Bose-Hubbard model
\begin{align}\label{eq:BHMapp}
\hat{\mathcal{H}}_{\rm BH} &= - J \sum\limits_{i, j=i\pm 1} \left(\hat{a}_{i}^{\dagger}\hat{a}_{j} + \hat{b}_{i}^{\dagger}\hat{b}_{j}\right) + \frac{U_{aa}}{2}\sum\limits_{i} \hat{n}^a_i (\hat{n}^a_i -1) \nonumber\\
& + \frac{U_{bb}}{2}\sum\limits_{i} \hat{n}^b_i (\hat{n}^b_i -1) 
+ U_{ab} \sum \limits_{i} \hat{n}^a_i\hat{n}^b_i ,
\end{align}
where $J=J(1)$, $U_{\sigma \sigma'}=U^{\sigma \sigma'}_{0,0,0,0}$ and $\hat{n}^a_i=\hat{a}_{i}^{\dagger}\hat{a}_{i}$, $\hat{n}^b_i=\hat{b}_{i}^{\dagger}\hat{b}_{i}$. In the shallow lattice however, for $V_0 \ll 1$, all the terms in (\ref{eq:ham}) have to be taken into account.

\section{Numerical calculation of the Bloch functions}\label{ap:bloch}

The eigenfunctions of the single-particle Hamiltonian~(\ref{eq:single_particle_hamiltonian}), 
\begin{equation}
\hat{h}(t) = -\frac{\hbar^2}{2 m} \frac{d^2}{dx^2} + V(x),
\end{equation}
are Bloch functions $\psi_{l, q}(x)$.
They can be calculated numerically in a quite straightforward way. One starts with the representation
\begin{equation}\label{eq:app:Bloch}
\psi_{l, q}(x)= e^{i q x}u_{l, q}(x) 
\end{equation}
where $q$ is the quasimomentum, here an integer multiple of $2\pi/(M d)$, $l$ denotes the energy band and $x$ is a position. Since the functions $u_{ q}(x)$ are periodic with the same periodicity of the lattice, they can be written as a discrete Fourier sum
\begin{equation}\label{eq:q:ulq}
u_{l, q}(x)=\sum_{j} a_{j} e^{i 2j k x},
\end{equation}
while the potential $V(x)=V_0 \sin^2(k x)$ as
\begin{equation}
V(x)=\frac{V_0}{2}\left(  1 - \cos(kx) \right)=\frac{V_0}{2}-\frac{V_0}{4}\left( e^{i kx} + e^{-i kx} \right).
\end{equation}
Then, the stationary Shr\"odinger equation $\hat{h} \psi_{l, q}(x)=E \psi_{l, q}(x)$, after multiplying both sides by $e^{ −iqx}$ and performing the integral $\int dx $, gives the equation for coefficients of decomposition in the Fourier basis
\begin{equation}
\left[ \left( 2 j +\frac{q}{k} \right)^2 + \frac{V_0}{2E_R} \right] a_j -\frac{V_0}{4 E_R}\left( a_{j+1} + a_{j-1} \right)=\frac{E}{E_R}a_j.
\end{equation}
This is a standard eigenvalue problem and a numerical solution of the polynomial equation $\det(M- E \mathbb{1}) = 0$ with
\begin{equation} 
\frac{M_{j,j'}}{E_R} = \left\{ 
\begin{array}{ll}
\left( 2j + \frac{q}{k} \right)^2 + \frac{V_0}{2 E_R}, & \mbox{if $j=j'$};\\
-\frac{V_0}{4 E_R}, & \mbox{if $|j-j'|=1$};\\
0, & \mbox{otherwise},
\end{array} 
\right. 
\end{equation} 
provides $(2j_{\rm max} + 1)$ eigenvalues of the matrix $M$. These eigenvalues are
labeled by the index $l = 1, 2, . . . , 2 j_{\rm max} + 1$ (called band index) and denoted as $E_{l,q}$. The components of an eigenvector corresponding to given eigenvalue $E_{l,q}$ are the coefficients $a_j$ of (\ref{eq:q:ulq}). Having them, one needs to evaluate the sum in (\ref{eq:q:ulq}) to extract the Bloch function (\ref{eq:app:Bloch}).

\section{Properties of tunneling and interaction terms}\label{app:JandU}

\begin{figure}[bt!]
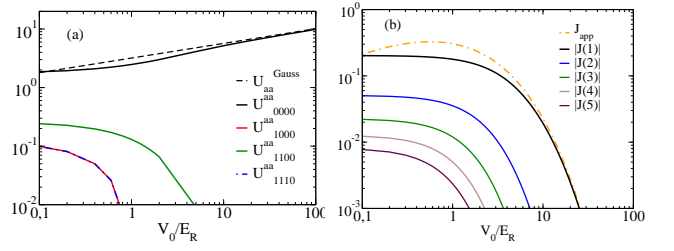

\centering
\includegraphics[width=.49\linewidth]{Fig10a.eps}
\includegraphics[width=.475\linewidth]{Fig10b.eps}
\caption{(a) Interaction terms $U^{aa}_{i,j,k,l}/\left[ a_{\sigma \sigma'}d/(L_x L_y) \right]$ defined in Eq. (\ref{eq:u_definition}) calculated numerically using the Wannier function for the values of $i,j,k,l$ as indicated in legend (solid lines). The black dashed line shows the result of Gaussian approximation for $U_{0000}/\left[ a_{\sigma \sigma'}d/(L_x L_y) \right]$~(\ref{eq:GaussU}).
(b) Tunneling terms in units of the recoil energy $E_R$ calculated exactly from the energy spectrum of the single-particle Hamiltonian in the case of different $i-j$ sites distances as indicated in legend. The approximated result for the nearest neighbor~(\ref{eq:GaussJ}) is shown by the orange dot-dashed line. }
\label{fig:app:fig1}
\end{figure}

In the limit $V_0/E_R \gg 1$ the Wannier functions can be approximated by Gaussian functions whose width is set by the frequency associated to the each lattice site minimum~\cite{RevModPhys.80.885}. Therefore, $w(x)\approx \left( \frac{k^2}{\pi} \right)^{1/4} \left( \frac{V_{0}}{E_R} \right)^{1/8} e^{-\sqrt{V_{0}} k^2 x^2/2}$. This is a fairly good approximation to obtain the interaction coefficient~:
\begin{equation}
   \frac{U^{\rm Gauss}_{\sigma \sigma'} (t) }{E_R} \approx \sqrt{\frac{32}{\pi}} \frac{a_{\sigma \sigma'} d}{L_x L_y} \left(\frac{V_{0}}{E_R}\right)^{1/4} . \label{eq:GaussU}
\end{equation}
In Fig.~\ref{fig:app:fig1}(a) we show the interacting terms calculated exactly from (\ref{eq:u_definition}) and compare them to the approximated formula (\ref{eq:GaussU}). One can easily see that the interaction terms beyond the terms involving nearest neighbors can be neglected as compared to the latter ones, when $V_0/E_R> 1$ as expected~\cite{RevModPhys.80.885}.

In the case of the hopping parameters $J(i-j)$, the Gaussian approximation gives a relative error growing with the lattice height~\cite{PhysRevA.79.053623}. The hopping matrix element is rather conveniently approximated by the width of the lowest band in the one-dimensional Mathieu equation~\cite{Zwerger_2003}, yielding to
\begin{equation}
    \frac{J_{\rm app}}{E_R} \approx \frac{4}{\sqrt{\pi}} \left(\frac{V_{0,x}}{E_R}\right)^{3/4} e^{-2\sqrt{\frac{V_{0,x}}{E_R}}}, \label{eq:GaussJ}
\end{equation}
for nearest neighbors, i.e. $i-j=1$. 
In Fig.~\ref{fig:app:fig1}(b) we present the tunneling terms calculated exactly and using the above formula. Indeed, as long as $V_0/E_R>1$ the leading nearest neighbor term dominates.

\section{Evolution for $N=2$: analytical results}\label{app:Neq2}

In below we present the semi-analytical method used to describe the dynamics of $N=M=2$ system.

Let us consider the Bose-Hubbard Hamiltonian (\ref{eq:BHM}) in the Fock state basis, which for $N=M=2$ is ordered and named as follows
\begin{align}\label{eq:Fsbasis}
    \ket{1}&=\ket{20,00}, \, 
    \ket{2}=\ket{11,00}, \,
    \ket{3}=\ket{02,00},\nonumber \\
    \ket{4}&=\ket{10,10},\,
    \ket{5}=\ket{10,01},\,
    \ket{6}=\ket{01,10},\,
    \ket{7}=\ket{01,01},\nonumber \\
    \ket{8}&=\ket{00,20},\,
    \ket{9}=\ket{00,11},\,
    \ket{10}=\ket{00,02}.
\end{align}
For example $\hat{a}^\dagger_2 \hat{a}_1 \ket{1}=\sqrt{2} \ket{2}$.
The Hamiltonian matrix in the above basis has block diagonal form 
$\hat{{\cal H}}_{BH}~=~\text{diag}\{ \hat{{\cal H}}_\text{aa}, \hat{{\cal H}}_\text{ab}, \hat{{\cal H}}_\text{bb} \}$
where matrix representation of $\hat{{\cal H}}_{aa}$ and $\hat{{\cal H}}_{bb}$ in states $\{|1\rangle, |2\rangle, |3\rangle \}$ and $\{|8\rangle, |9\rangle, |10\rangle \}$, respectively, has form 
\begin{equation}\label{eq:app:Hss}
\hat{{\cal H}}_{aa} = \hat{{\cal H}}_{bb}  = 
\left(
\begin{array}{ccc}
 U_{\sigma \sigma} & -2 \sqrt{2} J & 0 \\
 -2 \sqrt{2} J & 0 & -2 \sqrt{2} J \\
 0 & -2 \sqrt{2} J & U_{\sigma \sigma} \\
\end{array}
\right),
\end{equation}
and $\hat{{\cal H}}_{ab}$ representation in states $\{|4\rangle, |5\rangle, |6\rangle, |7\rangle \}$ reads
\begin{equation}
\hat{{\cal H}}_{ab}=\left(
\begin{array}{cccc}
 U_{ab} & -2 J & -2 J & 0 \\
 -2 J & 0 & 0 & -2 J  \\
 -2 J & 0 & 0 & -2 J \\
 0 & -2 J & -2 J & U_{ab} \\
\end{array}
\right).
\end{equation}

In order to calculate the initial spin-coherent state, let us first consider the ground state when all atoms are in the $a$ state, and then rotate it through $\pi/2$ around the $y$-axis of the Bloch sphere. 
When all atoms are in the component $a$ then it is enough to consider the $3\times3$ matrix (\ref{eq:app:Hss}). Eigenvalues of that matrix are 
$E_1=\Omega_{aa-}/2$, $E_2=U_{aa}$, $E_3=\Omega_{aa+}/2$,
where $\Omega_{aa\pm}=U_{aa}\pm \sqrt{64J^2+U_{aa}^2}$, with the corresponding states:
\begin{align}
    |\Psi_1\rangle &= \frac{1}{\mathcal{N}_1}\left( |1\rangle + \frac{\Omega_{aa+}}{4\sqrt{2}J} |2\rangle +  |3\rangle\right) \\
    |\Psi_2\rangle &= \frac{1}{\sqrt{2}}\left( -|1\rangle + |3\rangle\right) \\
    |\Psi_3\rangle &= \frac{1}{\mathcal{N}_3}\left( |1\rangle + \frac{\Omega_{aa-}}{4\sqrt{2}J} |2\rangle +  |3\rangle\right),
\end{align}
where $\mathcal{N}_1^2=2+\Omega_{aa+}^2/(32J^2)$ and $\mathcal{N}_3^2=2+\Omega_{aa-}^2/(32J^2)$.
The ground state before rotation is therefore
\begin{equation}
    \ket{GS^-}=\ket{\Psi_1}=\sum_{w=1}^{10} c^{-}_w \ket{w}
\end{equation}
with $\vec{c}^{-}=(1,\frac{\Omega_{aa=}}{4\sqrt{2}J},1,0,0,0,0,0,0,0)^{T}/\mathcal{N}_1$, as $E_1$ has the lowest value.

The $\pi/2$ pulse $e^{- i\hat{S}_y \pi/2}$ in the basis we consider is
{\footnotesize 
\begin{equation}
    \left(
\begin{array}{cccccccccc}
 \frac{1}{2} & 0 & 0 & -\frac{1}{\sqrt{2}} & 0 & 0 & 0 & \frac{1}{2} & 0 & 0 \\
 0 & \frac{1}{2} & 0 & 0 & -\frac{1}{2} & -\frac{1}{2} & 0 & 0 & \frac{1}{2} & 0 \\
 0 & 0 & \frac{1}{2} & 0 & 0 & 0 & -\frac{1}{\sqrt{2}} & 0 & 0 & \frac{1}{2} \\
 \frac{1}{\sqrt{2}} & 0 & 0 & 0 & 0 & 0 & 0 & -\frac{1}{\sqrt{2}} & 0 & 0 \\
 0 & \frac{1}{2} & 0 & 0 & \frac{1}{2} & -\frac{1}{2} & 0 & 0 & -\frac{1}{2} & 0 \\
 0 & \frac{1}{2} & 0 & 0 & -\frac{1}{2} & \frac{1}{2} & 0 & 0 & -\frac{1}{2} & 0 \\
 0 & 0 & \frac{1}{\sqrt{2}} & 0 & 0 & 0 & 0 & 0 & 0 & -\frac{1}{\sqrt{2}} \\
 \frac{1}{2} & 0 & 0 & \frac{1}{\sqrt{2}} & 0 & 0 & 0 & \frac{1}{2} & 0 & 0 \\
 0 & \frac{1}{2} & 0 & 0 & \frac{1}{2} & \frac{1}{2} & 0 & 0 & \frac{1}{2} & 0 \\
 0 & 0 & \frac{1}{2} & 0 & 0 & 0 & \frac{1}{\sqrt{2}} & 0 & 0 & \frac{1}{2}
\end{array}
\right),
\end{equation}
}
leading to the spin coherent state
\begin{equation}\label{eq:Sh}
    \ket{\Psi(0)}=e^{- i\hat{S}_y \pi/2} \ket{GS^-} = \sum_w c_w(0) \ket{w},
\end{equation}
with
\begin{equation}
    \vec{c}_w(0)=(2\sqrt{2}, \frac{a}{2}, 2\sqrt{2},4,\frac{a}{2},\frac{a}{2},4,2\sqrt{2},\frac{a}{2},2\sqrt{2} )^{T}\frac{1}{\sqrt{8^2 + a^2}}
\end{equation}
and $a=\frac{\Omega_{aa+}}{J}$.
A solution of the Schr\"odinger equation $i \hbar \partial_t \ket{\Psi(t)} = H \ket{\Psi(t)}$ reads
\begin{equation}\label{eq:Shro}
    \ket{\Psi(t)}=e^{-i t H/\hbar} \ket{\Psi(0)},
\end{equation}
when the coefficients in the Bose-Hubbard Hamiltonian are time-independent, or has to be calculated numerically in the time-dependent case.

The time-independent case (\ref{eq:Shro}) can be solved analytically, however the general expressions are quite complex. Here, we give a result for the symmetric case when $U_{aa}=U_{bb}$, which is
\begin{equation}
    \ket{\Psi(t)} = \sum_w c_w(t) |w\rangle,
\end{equation}
with
\begin{align}
    &c_1(t)=\frac{2Je^{-it\Omega_{aa-}/2}}{\sqrt{64J^2+U_{aa}\Omega_{aa+}}},\nonumber \\
    &c_2(t)=\frac{\sqrt{\Omega_{aa+}}e^{-it\Omega_{aa-}/2}}{2\sqrt{2\omega_{aa}}}, \nonumber \\
    &c_4(t)=\frac{2Je^{-it\Omega_{ab+}/2}
    \left[ \Omega_{ab+} - \Omega_{aa+} + e^{i t \omega_{ab}}(\Omega_{aa+} - \Omega_{ab-})\right]}{\sqrt{\omega^2_{ab}(64J^2 + \Omega_{aa+}^2)}} ,\nonumber \\
    &c_5(t)=\frac{e^{-itU_{ab}/2}}{2\sqrt{2 \omega^2_{ab} (64J^2 + U_{aa}\Omega_{aa+})}} \times \nonumber \\
    &\times
    \left[ \Omega_{aa+} \omega_{ab}{\rm cos}(t\omega_{ab}/2)
    +i (64J^2+\Omega_{aa+}U_{ab}){\rm sin}(t\omega_{ab}/2) \right] , \label{eq:c1-c5}
\end{align}
and $c_3(t)=c_8(t)=c_{10}(t)=c_1(t)$, $c_6(t)=c_5(t)$, $c_7(t)=c_4(t)$, $c_9(t)=c_2(t)$.
In the above expressions $\omega_{\sigma\sigma'} = \sqrt{64J^2 + U_{\sigma \sigma'}}$ and $\Omega_{\sigma \sigma' \pm}=U_{\sigma \sigma'} \pm \omega_{\sigma \sigma'}$. The reason why coefficients $c_6-c_{10} $ are expressed in terms of $c_1-c_5$ is the symmetry in respect to exchange of $a$ and $b$, forced by equality of interaction coefficients $U_{aa}=U_{bb}$. Otherwise, they are different and have complex analytical forms. Note, in the limit $J\to 0$ coefficients $c_2, \, c_5, \, c_6, \, c_9$ are nonzero while the remaining ones $c_1, c_3, c_4, c_7, c_8, c_{10}$ tend to zero. Therefore, in the Mott phase only the Fock states with one atom per lattice site contributes to the nonzero values of an observable. The same holds in the non symmetric case as well.

\bibliography{bibliography}

\end{document}